\documentclass[english,floatfix,twocolumn,superscriptaddress,nofootinbib,showpacs,prd]{revtex4}
\usepackage[paperwidth=210mm,paperheight=297mm,centering,hmargin=2cm,vmargin=2cm]{geometry}
\usepackage[utf8]{inputenc}
\usepackage[T1]{fontenc}
\usepackage{lmodern}
\usepackage{amsmath}
\usepackage{amssymb}
\usepackage{graphicx}
\usepackage{subcaption}
\usepackage{esint}
\usepackage{dcolumn}
\usepackage{babel}
\usepackage{xcolor} 
\usepackage{csquotes}
\usepackage{marvosym}
\usepackage{hyperref}
\usepackage[overload]{textcase}
\usepackage[symbol,hang,flushmargin]{footmisc}

\DefineFNsymbolsTM{otherfnsymbols}{%
  \Bat	\circ
  \Yinyang   \mathsection
  \Biohazard    \dagger
  \Lightning \ddagger
  \Radioactivity \mathsection
}%

\setfnsymbol{otherfnsymbols}

\newenvironment{seqn}{\equation\aligned}{\endaligned\endequation}

\newcommand{\be}{\begin{seqn}}
\newcommand{\ee}{\end{seqn}}

\newcommand{\bea}{\begin{eqnarray}}
\newcommand{\eea}{\end{eqnarray}}

\newenvironment{arabicfootnotes}
  {\par\edef\savedfootnotenumber{\number\value{footnote}}
   
   \setcounter{footnote}{0}}
  {\par\setcounter{footnote}{\savedfootnotenumber}}

\begin{document}
%%%%%%%%%
%
%
%
%
%%%%%%%%%%%%%%%%%%%%%%%%%%%%%%%%%%%%%%%%%%%%
\title{Linearized Stability of Bardeen de-Sitter Thin-Shell Wormholes}

\author{Hassan~Alshal}
\email{halshal@sci.cu.edu.eg}
\affiliation{Department of Physics, Faculty of Science, Cairo University, Giza, 12613, Egypt.}
\affiliation{Department of Physics, University of Miami, Coral Gables, FL 33146, USA.}

\begin{abstract}

\begin{center}
\textbf{ABSTRACT}
\end{center}

\par\noindent
A thin-shell wormhole is crafted by the cut-and-paste method of two Bardeen de-Sitter black holes using Darmois-Israel formalism.  Energy conditions are considered for different values of magnetic charge while both mass and cosmological constant are fixed. The attractive and repulsive characteristics of the throat of the thin-shell wormhole are also examined through the radial acceleration. Dynamics and stability of the wormhole are studied around the static solutions of the linearized radial perturbations at the throat of the wormhole. The regions of stability are determined by checking out the condition of concavity of the potential as a function in the throat radius for different values of magnetic charges.
\end{abstract}
\maketitle
\begin{arabicfootnotes}
%
%
%
%
%%%%%%%%%%%
\section{Introduction}
\par\noindent
The first wormhole solution was discovered by Ludwig Flamm \cite{Flamm:1916er}. It was rediscovered as the \textit{Einstein-Rosen bridge} \cite{Einstein:1935tc} while Einstein and Rosen were trying to develop a non-Boscovichian, i.e., singularity-free, atomic model of gravity and electromagnetism. Later, Wheeler developed a theory about \textit{geons} \cite{Wheeler:1955zz}, topologically unstable gravitoelectromagnetic quasi-solitons that can connect widely separated spacetime regions. Misner and Wheeler tried to develop the theory of geons into a geometrical unified classical theory \cite{Misner:1957mt}. In Misner and Wheeler project the \textit{wormhole} term was coined.\\
\par\noindent
Between the development of geons and the rejuvenation of Morris and Thorne \textit{traversable} wormholes \cite{Morris:1988cz}, Ellis studied the flow of ``substantial ether'' through a \textit{drainhole} \cite{Ellis:1973yv}. Also Bronnikov analyzed tunnel-like solutions \cite{Bronnikov:1973fh}, which are considered the precursors to the studies of wormholes in modified theories of gravity \cite{Lobo:2017oab}.  Geons reappear again in \textit{galileon} theory as a scalar-tensor theory \cite{Curtright:2012ph}. Even in Euclidean space, Ellis variant \textit{p-norm} drainholes can be used as \textit{pedagogical} examples to study electrostatics \cite{Curtright:2018qif, Alshal:2018hsg, Alshal:2018srh, Alshal:2019smy}. A recent study discusses erudite and deep-seated examples, in both flat space and curved spacetime, in which the objects behavior around wormholes are analyzed in terms of scalar, \emph{electromagnetic}, and gravitational fields \cite{Dai:2019mse}. More on wormholes can be found in Ref. \cite{Visser:1995cc}.\\
\par\noindent
To find a wormhole solution to field equation, one can choose some equations of state such as phantom energy \cite{Sushkov:2005kj, Lobo:2005yv}, Chaplygin gas \cite{Lobo:2005vc}, and/or quintessence \cite{Lobo:2006ue}. Then, rotating spacetimes \cite{Teo:1998dp}, evolving wormholes \cite{Kar:1995ss}, thin-shell spacetimes \cite{Poisson:1995sv}, dust shell wormholes \cite{Lobo:2004rp} and/or Casimir wormholes \cite{Garattini:2019ivd} can be implemented to the field equations to ``ameliorate'' the violation of energy conditions associated with the \textit{flaring-out} condition, which is necessary for the field equations to have wormhole solutions. There are numerous studies that consider different black holes creating thin-shell wormholes in de-Sitter and anti-de-Sitter spacetimes \cite{Kuhfittig:2010pb, Rahaman:2011yh, Sharif:2013xta, Sharif:2014ria, Eid:2015pja, Ovgun:2017jzt}. Stability of these thin-shell wormholes are examined too \cite{Ishak:2001az, Lobo:2005zu, Eiroa:2007qz, Eiroa:2008ky, Lemos:2008aj, Dias:2010uh, Eiroa:2011nd, Sharif:2013nka, Mazharimousavi:2014gpa, Lobo:2015lbc, Eid:2016axb, Ovgun:2017jip, Amirabi:2017buh, HabibMazharimousavi:2017zlc, Eiroa:2017nar, Tsukamoto:2018lsg, Forghani:2019wgt}. Also thin-shell wormholes can be obtained from regular black holes \cite{Halilsoy:2013iza, Sharif:2016gyb}, which is the general theme of this letter.\\
\par\noindent
In this letter we construct Bardeen de-Sitter thin-shell wormholes. The Bardeen black hole \cite{Bardeen:1968bk} is an interesting \textit{regular} black  hole, i.e., with no \textit{geometric} singularity. Bardeen black hole can be discerned as a quantum-corrected Schwarzschild black hole \cite{Maluf:2018ksj} by applying the \textit{generalized uncertainty principle} \cite{Amati:1988tn, Garay:1994en, Scardigli:1999jh, Adler:2001vs, Ali:2009zq, Faizal:2017dlb, Vagenas:2018pez, Vagenas:2019wzd, Vagenas:2019rai} to the black hole and studying the consequent effects on the corresponding thermodynamics. Singularity-free black hole comes with special feature that is such black hole does not continue collapsing until it completely evaporates \cite{Hayward:2005gi}. Instead, the outer horizon of a regular black hole is shrinking meanwhile the inner horizon is growing until the two horizons meet \cite{Hayward:2005ny,Nester:2007ps}. Therefore, employing a regular black hole in thin-shell wormhole construction comes with the advantage of having viable spacetime, especially in particle collisions \cite{Pradhan:2014oaa}, despite the raised questions on the stability of the inner horizon and the evaporation timescale \cite{Carballo-Rubio:2018pmi}. Bardeen black hole can be used in de-Sitter background (BdS) \cite{Fernando:2016ksb}. The BdS black hole anti-evaporation scenario has been also studied \cite{Singh:2017}. The BdS solution in arbitrary dimensions and the corresponding thermodynamics for each dimension are also considered \cite{Ali:2018boy}. One can implement electric charge to construct thin-shell wormholes from regular ABG black holes \cite{Sharif:2016gyb}. Magnetic monopoles are hypothetically created in Beyond Standard Model theories. Despite they have not been discovered yet in nature, the anti-evaporation scenario makes the Bardeen black holes more stable compared with the ABG black holes. This is because ABG black holes quickly discharge through Hawking radiation and pair creations when they are close to the quantum Planck scale \cite{Ansoldi:2006vg,Nicolini:2017hnu}. The nonlinear electrodynamics associated with such regular black holes could keep the weak energy condition satisfied \cite{Balart:2014jia}.\\
\par\noindent
In section ($\mathbf{II}$), we use Visser's technique of cut-and-paste \cite{Visser:1989kh, Visser:1989kg}, together with Darmois-Israel formalism \cite{Mansouri:1996ps}, to connect two BdS regions of spacetime through a thin shell. Cut-and-paste method ensures utilizing diminutive amount of exotic matter, hence a traveler through such wormholes can avoid the regions where the exotic matter is \cite{Visser:1989kh}. The exotic matter is confined at the thin-shell regions similar to the matter of ABG thin-shell wormholes \cite{Sharif:2016gyb}. We also study the components of the stress-energy-momentum surface tensor using the extrinsic curvature. We comment on the violation of energy conditions, because of the exotic matter at the wormhole throat, in terms of stress components. The jump in the extrinsic curvature is considered upon passing through charged shells \cite{Kuchar:1968}. From the discontinuity in the extrinsic curvature, we can compare the effect of magnetic charges on our chosen non-vacuum spacetime to the vacuum spacetime in Ref. \cite{Lobo:2003xd}. We also calculate the attraction and repulsion nature of the wormhole throat in terms of the acceleration.\\
\par\noindent
In section ($\mathbf{III}$), we analyze the linear stability of BdS thin-shell wormhole by studying the concavity condition on the ``speed of sound'' as a function in BdS parameters: the mass, the magnetic monopoles and the cosmological constant. And we see the change in stability regions upon varying the amount of magnetic monopoles while both mass and cosmological constant are fixed.\\
\par\noindent
In section ($\mathbf{IV}$) we summarize and comment on the results of the previous two sections. 
%%%%%%%%%%%%
%
%
%
%
%%%%%%%%%%%
\section{Visser's Cut-and-Paste Technique and the Darmois-Israel formalism}
%%%%%%%%%%%%
%
%
\par \noindent
The BdS black hole is constructed \cite{Fernando:2016ksb} starting with the metric
\be\label{eq.1}
ds^2_{\text{BdS}}=&-\left(1-\frac{2 m(r)}{r}\right)dt^2+\left(1-\frac{2 m(r)}{r}\right)^{-1} dr^2\\
&+r^2d\theta^2+r^2 \sin^2(\theta)d\phi^2~.
\ee
The field equations are derived from the action \cite{AyonBeato:2000zs}:
\be
\mathcal{A}=\int d^4 x \sqrt{-g} \left[\frac{R-2\Lambda}{16\pi} -\frac{1}{4\pi}\frac{3M}{|\mu|^3}\left(\frac{\sqrt{2\mu^2 F}}{1+\sqrt{2\mu^2 F}}\right)^{\frac{5}{2}}\right],
\ee
where $\mu$ is the magnetic monopole charge, M is the mass of the black hole, and $\displaystyle{F=\frac{1}{4}F^{\mu \nu}F_{\mu \nu}}$ is the non-interacting part of Lagrangian density for a classical electrodynamics field tensor $F_{\mu\nu}$.\\
Therefore, eq.\eqref{eq.1} becomes:
\be
ds^2_{\text{BdS}}=-f(r)dt^2+f(r)^{-1} dr^2+r^2d\theta^2+r^2 \sin^2(\theta)d\phi^2,
\ee
where

\be \label{eq.4}
f(r)= 1-\frac{2Mr^2}{(r^2+\mu^2)^{3/2}}-\frac{\Lambda}{3}r^2~.
\ee
And by finding the roots of $f(r)=0$, or the roots of the \textit{decic} polynomial

\be
&r^{10}\Lambda^2 +r^8 \left(3 \Lambda ^2 \mu^2 - 6 \Lambda \right)+r^6 \left(3 \Lambda ^2 \mu^4 - 18 \Lambda \mu^2 + 9\right)\\
&+r^4 \left(\Lambda ^2 \mu^6 - 18 \Lambda \mu^4 + 27 \mu^2 - 36 M^2\right)\\
&+r^2 \left(27 q^4-6 \Lambda \mu^6\right)+9 \mu^6=0,
\ee
one can determine the location of the inner, event ($r_h$) and cosmological ($r_c$) horizons of the BdS. However, we must avoid the combinations of $M, \mu$, and $\Lambda$ that lead to formation \textit{extreme} BdS \cite{Sharif:2014ria}\textemdash at which the event and cosmological horizons coincide by setting $f(r)=f'(r)=0$ \textemdash so the throat radius $a$ of the wormhole still exists as $r_h < a < r_c$.\\
\begin{figure}[h!]
\centering
%\begin{minipage}[t]{0.48\linewidth}
\includegraphics[width=\linewidth]{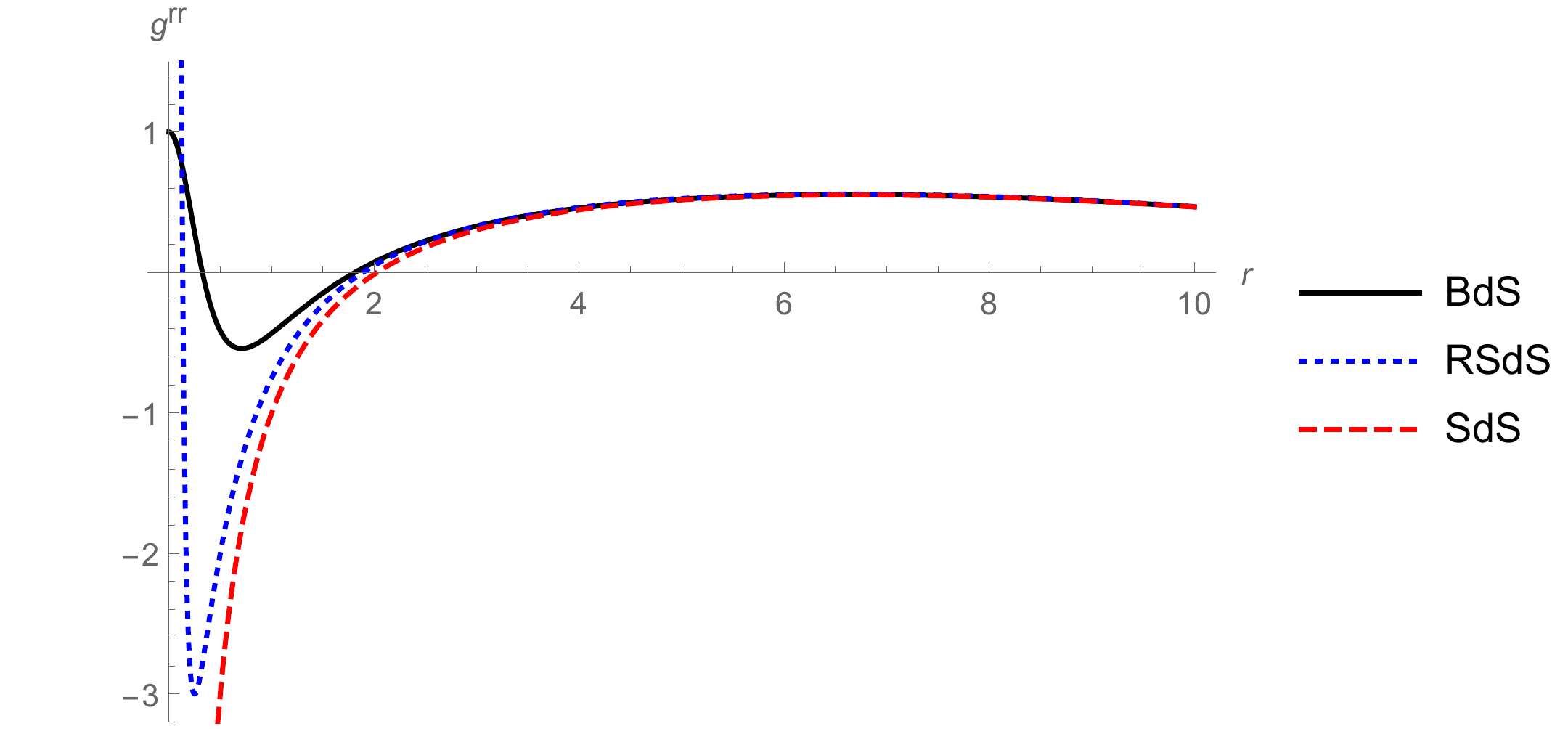}
\caption{The behavior of $g^{rr}$ metric component for Bardeen de-Sitter
(BdS: $M=1,\Lambda=0.01,\mu=0.5$),\\
Reissner-Nordstrom de-Sitter\\
(RNdS: $M=1,\Lambda=0.01,Q=0.5$),\\
and Schwarzschild de-Sitter (SdS: $M=1,\Lambda=0.01$).}\label{fig.1}
\end{figure}\\
Following the cut-and-paste technique \cite{Visser:1989kg, Lobo:2003xd}, one can easily construct a geodesically complete manifold $\displaystyle{\Gamma=\Gamma_+ \bigcup~\Gamma_-}$ by \textit{pasting} the region of timelike hypersurfaces, named a \textit{thin shell} $\displaystyle{\partial\Gamma=\partial\Gamma_+ \bigcup~\partial\Gamma_-}$~, where $\displaystyle{\partial\Gamma_{\pm} :=\left\lbrace r_{\pm}=a~|~a>r_h\right\rbrace}$, that bounds the bulk of two BdS. This follows after \textit{cutting} spacetime regions $\displaystyle{\Gamma_{\pm} :=\left\lbrace r_{\pm}\leq a~|~a>r_h\right\rbrace}$ inside the throat radius $a$.\\
Now we follow Darmois-Israel formalism \cite{Darmois:1927rt,Israel:1966rt} by defining the coordinates of $\Gamma$ as $x^{\mu}:=(t,r,\theta,\phi)$ and the coordinates of the shell $\partial\Gamma$ as $\zeta^i:=(\tau,\theta,\phi)$, where $\tau$ is the proper time that a comoving frame measures on the throat of the wormhole. The induced metric of the shell is:
\be
ds^2_{\partial\Gamma}= -d\tau^2 + r^2d\theta^2+r^2 \sin^2(\theta)d\phi^2,
\ee
where the parametric equation that relates $\Gamma$ to $\partial\Gamma$ is $r=a(\tau)$.\\
\par\noindent
In vacuum spacetime \cite{Lobo:2004id}, the interior solution $r_0$ is matched to the exterior one $a$ at the \emph{junction surface} $\partial\Gamma$, which is also known as thin-shell surface when the surface stress terms are present. The surface stresses are determined by the discontinuity in the extrinsic curvature $\mathcal{K}^i_j$. Junction surface restricts the exotic matter of the interior wormhole to a finite region\footnote{See figure 1 of Ref. \cite{Lobo:2004id}.}. To minimize the violation of the average null energy condition (ANEC), the wormhole should be constructed such that the exotic matter is restricted to the junction region $r_0<r<a$ with the limit $r_0\to a$ that turns the junction into thin-shell.\\
\par\noindent
For our BdS non-vacuum spacetime, the discontinuity in the extrinsic curvature is shown in terms of the surface stresses, just like the vacuum spacetime. However, the discontinuity is caused by the charges \cite{Kuchar:1968}. This discontinuity determines the jump in the electromagnetic field tensor $F_{\mu\nu}$ and, consequently, the jump in the electromagnetic stress-energy-momentum tensor across the thin-shell.\\
\par\noindent
We use the Gauss-Codazzi decomposition of spacetime such that it yields Israel's junction condition on $\Gamma$. The condition is described by the energy momentum tensor on the shell $\mathcal{S}^i_j=\text{diag}\left(-\sigma,p_{\theta},p_{\phi}\right)$ as
\be
\mathcal{S}^i_j=-\frac{1}{8\pi}\left(\left[\mathcal{K}^i_j\right] -\delta^i_j\mathcal{K}\right),
\ee  
where $\left[\mathcal{K}^i_j\right]=\mathcal{K}^{i~+}_j-\mathcal{K}^{i~-}_j$, and $\mathcal{K}=\left[\mathcal{K}^i_i\right]$.\\
We define the unit vectors $n^{\pm}_\mu$ normal to $\partial\Gamma$ as
\be \label{eq.8}
n^{\pm}_\mu=\pm\left(~\biggl\lvert g^{\alpha\beta}\frac{\partial f}{\partial x^{\alpha}}\frac{\partial f}{\partial x^{\beta}}\biggr\rvert^{-1/2}\frac{\partial f}{\partial x^{\mu}}\right)~.
\ee
And the extrinsic curvature, or the second fundamental form, is defined in terms of the unit vectors as
\be\label{eq.9}
\mathcal{K}^\pm_{ij}=-n_{\mu}\left(\frac{\partial^2 x^\mu}{\partial\zeta^i\zeta^j}+\Gamma^{\mu\pm}_{\nu\rho}~\frac{\partial x^\nu}{\partial \zeta^i}\frac{\partial x^\rho}{\partial \zeta^j}\right)
\ee
We substitute eq.\eqref{eq.4} in eq.\eqref{eq.8} to get:
\be \label{eq.10}
n^{\pm}_\mu=\biggl(\mp \dot{a},\pm\frac{\sqrt{\dot{a}^2+f(a)}}{f(a)},0,0\biggr)~.
\ee
Then, we substitute eq.\eqref{eq.10} in eq.\eqref{eq.9} to get the components of the extrinsic curvature as
\be
&\mathcal{K}^{\theta~\pm}_{\theta}=\mathcal{K}^{\phi~\pm}_{\phi} =\pm\frac{1}{a}\sqrt{1-\frac{2Ma^2}{(a^2+\mu^2)^{3/2}}-\frac{\Lambda}{3}a^2 + \dot{a}^2}~,\\
&\mathcal{K}^{\tau~\pm}_{\tau}=\pm\frac{\frac{6 M a^3}{\left(\mu ^2+a^2\right)^{5/2}}-\frac{4 M a}{\left(\mu ^2+a^2\right)^{3/2}}-\frac{2 \Lambda  a}{3}+\ddot{a}}{\sqrt{1-\frac{2Ma^2}{(a^2+\mu^2)^{3/2}}-\frac{\Lambda}{3}a^2 + \dot{a}^2}}~.
\ee
We use the the last results to define the surface stresses as
\be\label{eq.12}
\sigma=-\frac{1}{\pi}\mathcal{K}^\theta_\theta=-\frac{1}{2\pi a}\sqrt{1-\frac{2Ma^2}{(a^2+\mu^2)^{3/2}}-\frac{\Lambda}{3}a^2 + \dot{a}^2}~,
\ee
\be
p&=p_\theta=p_\phi=\frac{1}{8\pi}\left(\mathcal{K}^{\tau}_{\tau}+\mathcal{K}^{\theta}_{\theta}\right)\\
&=\frac{3}{8\pi a}\frac{\frac{1}{3}-\frac{2Ma^2}{(a^2+\mu^2)^{3/2}}-\frac{\Lambda}{3}a^2+\frac{2Ma^4}{(a^2+\mu^2)^{5/2}}+a\ddot{a}+\dot{a}^2}{\sqrt{1-\frac{2Ma^2}{(a^2+\mu^2)^{3/2}}-\frac{\Lambda}{3}a^2 + \dot{a}^2}}~.
\ee
And for the static configuration, i.e., $\dot{a}=\ddot{a}=0$, the surface stress become
\be\label{eq.14}
\sigma_0=-\frac{1}{2\pi a_0}\sqrt{1-\frac{2Ma^2_0}{(a^2_0+\mu^2)^{3/2}}-\frac{\Lambda}{3}a^2_0}~,
\ee
\be\label{eq.15}
p_0&=\frac{1}{8\pi}\left(\mathcal{K}^{\tau}_{\tau}+\mathcal{K}^{\theta}_{\theta}\right)\\
&=\frac{3}{8\pi a_0}\frac{1-\frac{2Ma^2_0}{(a^2_0+\mu^2)^{3/2}}-\frac{\Lambda}{3}a^2_0+\frac{2Ma^4_0}{(a^2_0+\mu^2)^{5/2}}}{\sqrt{1-\frac{2Ma^2_0}{(a^2_0+\mu^2)^{3/2}}-\frac{\Lambda}{3}a^2_0}}\\
&~~-\frac{1}{4\pi a_0}\frac{1}{\sqrt{1-\frac{2Ma^2_0}{(a^2_0+\mu^2)^{3/2}}-\frac{\Lambda}{3}a^2_0}}~.
\ee
From the last two equations, surface density $\sigma_0$ imposes the violation of the weak energy condition (WEC). Meanwhile, the null energy condition (NEC), $\sigma_0 +p_0>0$, can be maintained with no need to any \textit{exotic} effect from the combined mass and pressure of the matter as long as $\displaystyle{f(a_0)<\frac{6Ma^4_0}{(a^2_0+\mu^2)^{5/2}}}$~. And for the strong energy condition (SEC), $\sigma_0 +3p_0>0$, it is also maintained with $\displaystyle{f(a_0)<\frac{9 M a_0^4}{\left(\mu ^2+a_0^2\right)^{5/2}}-\frac{6 M a_0^2}{\left(\mu ^2+a_0^2\right)^{3/2}}-\Lambda a_0^2}$.
%Graphs
\begin{figure}[h!]
%\centering
\begin{minipage}[t]{\linewidth}
\captionsetup{justification=centering}
\includegraphics[width=\linewidth]{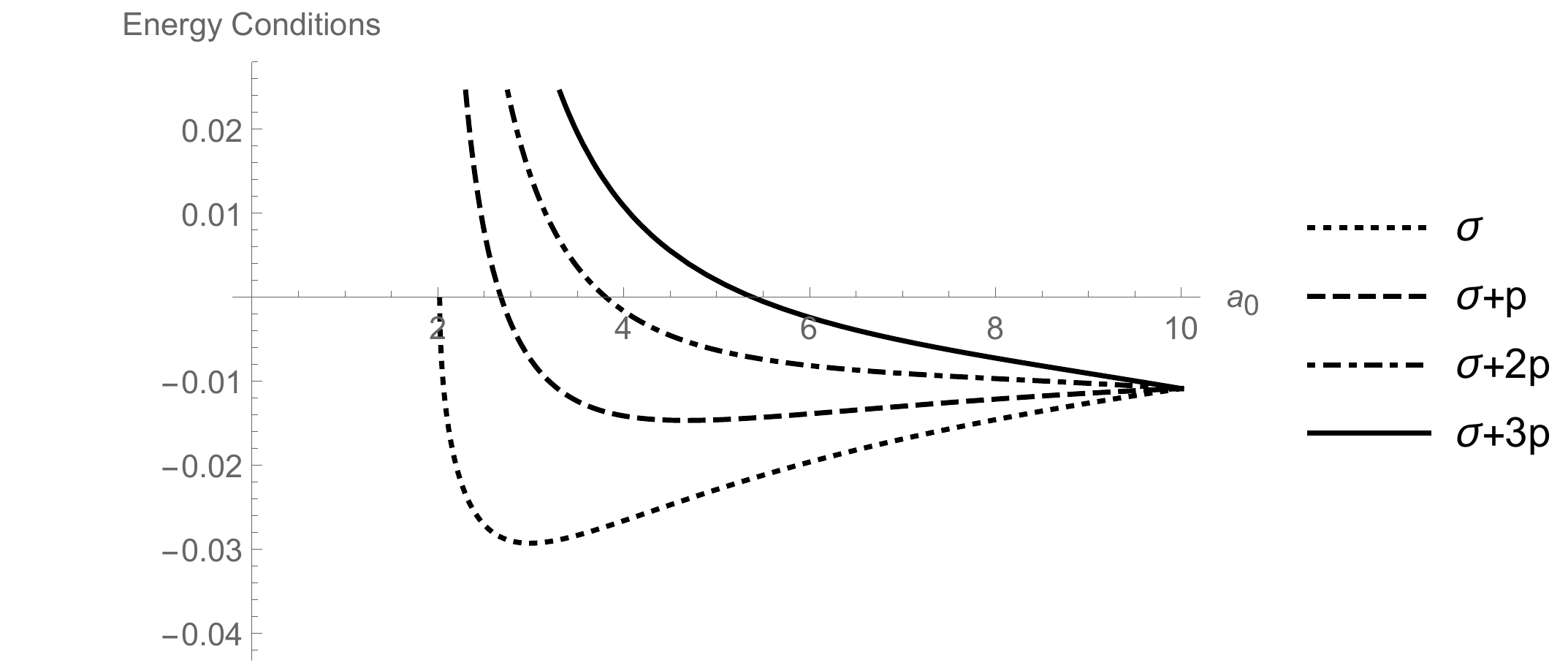}
\subcaption{ $\mu=0.1$.}
\end{minipage}\hfill

\begin{minipage}[t]{\linewidth}
\captionsetup{justification=centering}
\includegraphics[width=\linewidth]{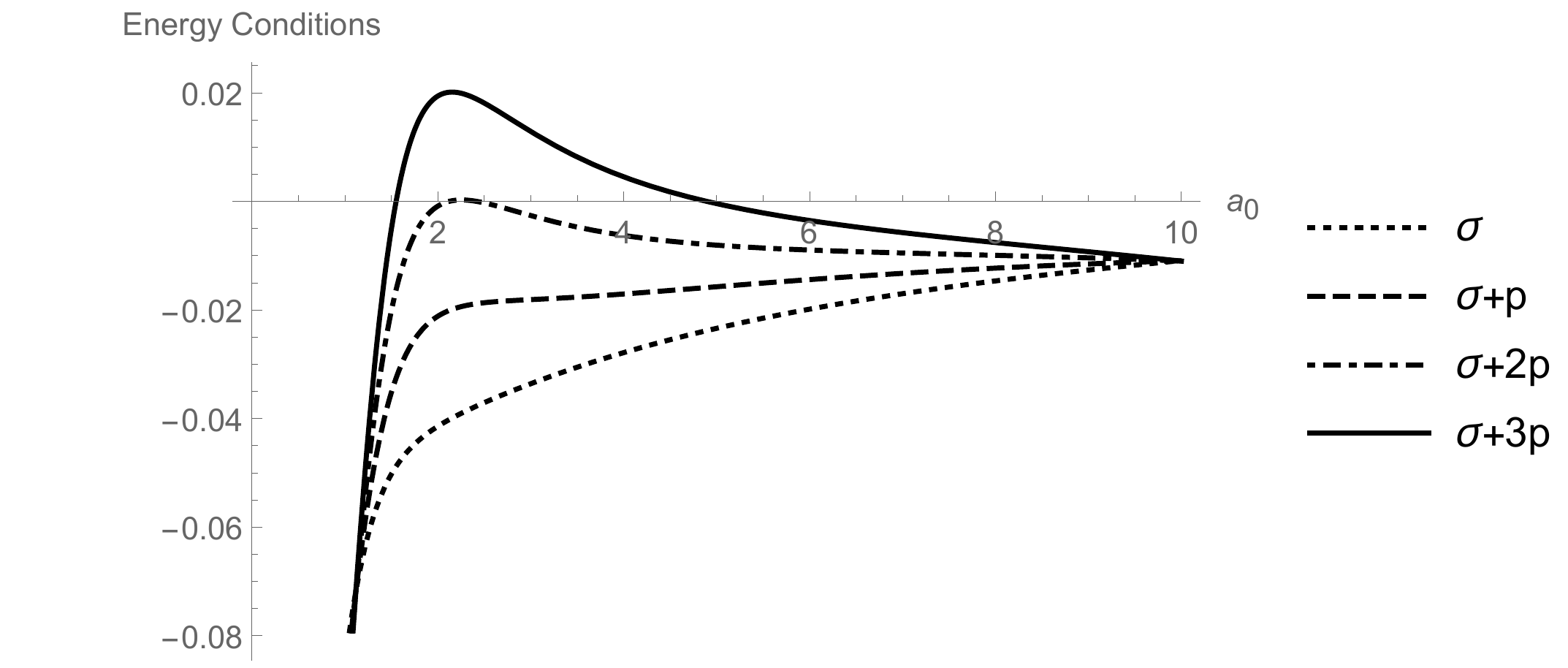}
\subcaption{ $\mu=1$.}
\end{minipage}
\caption{The energy conditions expressed in terms of $\sigma$ and $p$ vs. the throat radius $a_0$ with fixed $M=1$ and $\Lambda=0.01$, and different values of magnetic monopole: $\mu=0.1$ for figure.2.(a) and $\mu=1$ for figure.2.(b).}\label{fig.2}
\end{figure}
\par\noindent
For BdS black hole with no radial pressure, $p_r=0$, and a mass density that it localized at the throat $\rho=\sigma_0~\delta(r-a_0)$, the total amount of exotic matter necessary to keep the wormhole open is
\be
\Omega_{\sigma}&=\int^{2\pi}_0\int^{\pi}_0\int^{+\infty}_{-\infty}\sqrt{-g}~\sigma_0~\delta(r-a_0)~dr~d\theta~d\phi\\
&=-2a_0\sqrt{1-\frac{2Ma^2_0}{(a^2_0+\mu^2)^{3/2}}-\frac{\Lambda}{3}a^2_0}~.
\ee
We can examine the attractive and repulsive characters of the constructed thin-shell wormhole by studying the four-acceleration $a^\mu=u^\nu\nabla_\nu u^\mu$, where $u^\mu=(1/\sqrt{f(r)},0,0,0)$. The geodesic equation of a test particle is
\be
\frac{d^2 r}{d\tau^2}=-a^r~,
\ee
where the radial acceleration is given by
\be
a^r=\Gamma^r_{tt}\left(\frac{dt}{d\tau}\right)^2=\frac{Mr^3}{(r^2+\mu^2)^{5/2}}-\frac{\Lambda}{3}r -\frac{2M\mu^2 r}{(r^2+\mu^2)^{5/2}}~.
\ee
We notice that the wormhole has attractive or repulsive nature if $a^r>0$ or $a^r<0$ respectively.

%Graphs
\begin{figure}[h!]
%\centering
\begin{minipage}[t]{\linewidth}
\captionsetup{justification=centering}
\includegraphics[width=\linewidth]{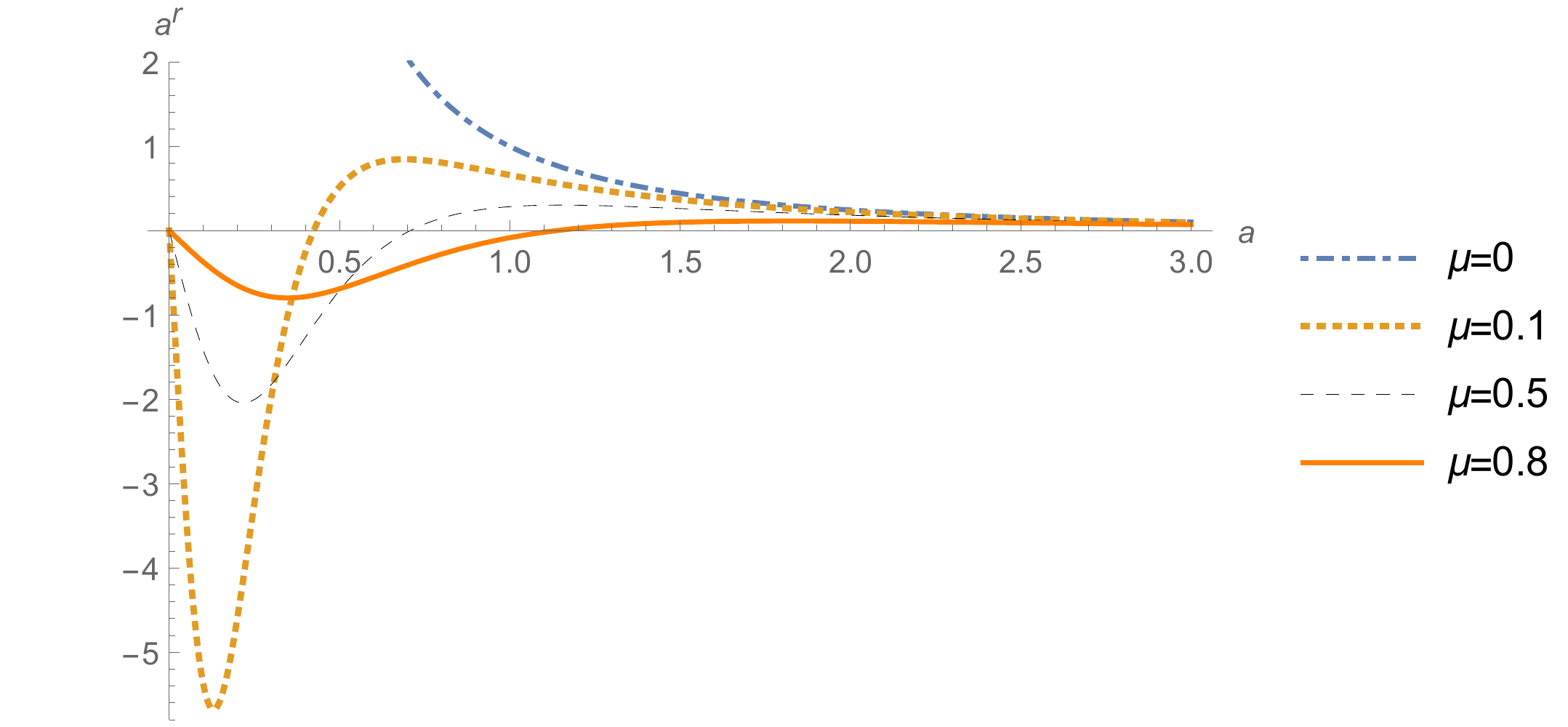}
\subcaption{Different attractive and repulsive behaviors of $a^r$ at small $a$ values for different $\mu$ values.}
\end{minipage}\hfill

\begin{minipage}[t]{\linewidth}
\captionsetup{justification=centering}
\includegraphics[width=\linewidth]{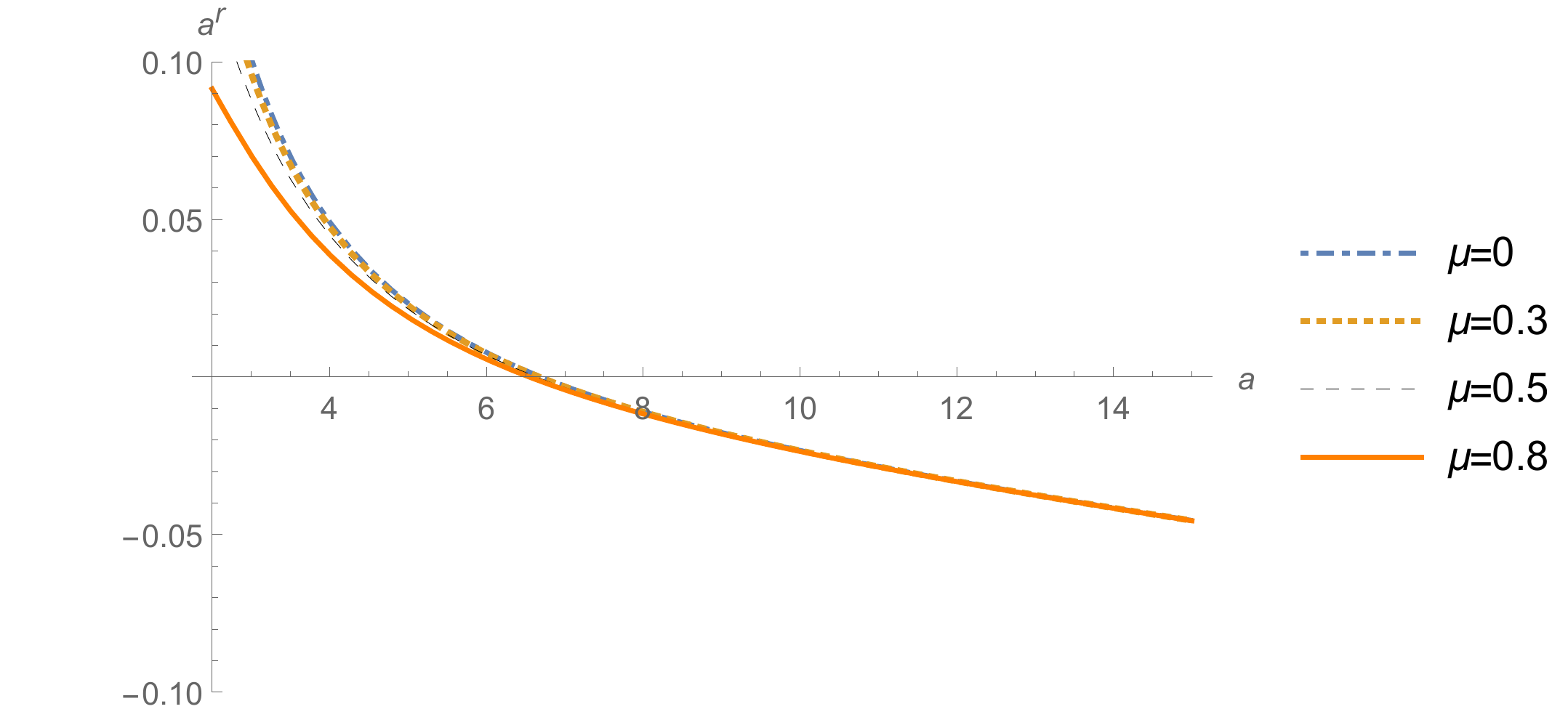}
\subcaption{Convergent behavior of $a^r$ at large $a$ values for different $\mu$ values. $a^r$ becomes repulsive at almost the same value of $a$ regardless the value of $\mu$.}
\end{minipage}
\caption{Attraction and repulsion in terms of acceleration $a^r$ vs. the throat radius $a$ with fixed $M=1$ and $\Lambda=0.01$, and different values of magnetic monopole $\mu$.}\label{fig.3}
\end{figure}
%
%
%
%
%%%%%%%%%%%%%%
\section{Linearized Stability Analysis}
\par\noindent
The stability of the wormhole can be checked \cite{Lobo:2003xd} by performing linear perturbation about the static configuration $(a=a_0)$ for eq.\eqref{eq.14} and eq.\eqref{eq.15}, where for vacuum spacetime $\mu=0$. One can easily notice that differentiating eq.\eqref{eq.12} with respect to $\tau$ yields the continuity equation
\be
\frac{d(\sigma A)}{d\tau}+p\frac{dA}{d\tau}=0~,
\ee
which directly leads to
\be \label{eq.20}
\sigma'=-\frac{2}{a}(\sigma+p)~,
\ee
where $A=4\pi a^2$ is the area of the wormhole throat, $\sigma'=\dot{\sigma}/\dot{a}$, the dot means $d/d\tau$, and the prime means $d/da$.\\
If we rearrange eq.\eqref{eq.12}, we define a potential function
\be \label{eq.21}
V(a)=f(a)-4\pi^2 a^2 \sigma^2=-\dot{a}~.
\ee
Then we  substitute with eq.\eqref{eq.20} in the first derivative of eq.\eqref{eq.21} to get
\be\label{eq.22}
V'(a)=& \frac{6 M a^3}{\left(\mu ^2+a^2\right)^{5/2}}-\frac{4 M a}{\left(\mu ^2+a^2\right)^{3/2}}\\
&-\frac{2 \Lambda a}{3} +8\pi^2 a\sigma(\sigma+2p)~.
\ee
And for the second derivative of \eqref{eq.21}, we parameterize the pressure to be a function in the density $p:=p(\sigma)$ \cite{Poisson:1995sv}. Then we introduce a new parameter $\vartheta(\sigma)=dp/d\sigma$, which can be seen as the ``speed of sound''. And the second derivative of \eqref{eq.21} becomes
\be\label{eq.23}
V''(a)&=f''(a)-8\pi^2\left[2\sigma(\sigma+p)(1+2\vartheta)+(\sigma+2p)^2\right]\\
&=f''(a)+\Bigg[~~\frac{1}{a^2}\bigg(af'(a)-2f(a)\bigg)\bigg(1+2\vartheta\bigg)\\
&\qquad\qquad\quad~~-\frac{1}{2}\left(\frac{f'(a)}{f(a)}\right)^2\Bigg]~.
\ee
To linearize the model, we apply Taylor expansion to the potential function around the static point $a=a_0$ such that eq.\eqref{eq.21} becomes
\be
V(a)=&V(a_0)+(a-a_0)V'(a_0)\\
&+\frac{1}{2}(a-a_0)^2V''(a_0)+\mathcal{O}\left[(a-a_0)^3\right]~.
\ee
We use eq.\eqref{eq.14} and eq.\eqref{eq.15} to evaluate eq.\eqref{eq.21} and eq.\eqref{eq.22} at $a=a_0$. Therefore, we get $V(a_0)=V'(a_0)=0$. Meanwhile eq.\eqref{eq.23} becomes
\be\label{eq.25}
V''(a_0)=&\frac{30 a^2_0 M}{\left(a^2_0+\mu ^2\right)^{5/2}}-\frac{4 M}{\left(a^2_0+\mu ^2\right)^{3/2}}-\frac{30 a^4_0 M}{\left(a^2_0+\mu^2\right)^{7/2}}\\
&-\frac{2 \Lambda }{3}
-\frac{1}{a_0^2}(1+2\vartheta)\left(\frac{6 M a_0^2}{\left(\mu ^2+a_0^2\right)^{5/2}}-\frac{2}{a_0^2}\right)\\
&-\frac{1}{2}\left(\frac{-\frac{4 M a_0}{\left(\mu ^2+a_0^2\right)^{3/2}}+\frac{6 M a_0^3}{\left(\mu ^2+a_0^2\right)^{5/2}}-\frac{2 \Lambda a_0}{3}}{1-\frac{2 M a_0^2}{\left(\mu ^2+a_0^2\right)^{3/2}}-\frac{\Lambda a_0^2}{3}}\right)^2~.
\ee
Of course we can use $(1+2\vartheta)=(\sigma'+2p')/\sigma'$ to express $\vartheta$ in terms of the metric parameters $M,\mu,$ and $a$ . But we will not as we need to study the behavior of $\vartheta$ when the throat is stable.\\
\par\noindent
The concave down condition $V''(a_0)<0$ results in provoking either expansion or contraction of the throat when any small perturbation occurs. While the convex, or the concave up, condition $V(a_0)''>0$ stabilizes the throat with a local minimum of $V(a_0)$ at $a_0$. Therefore, we solve for $\vartheta_0$ at that local minimum to get
\begin{widetext}
\be
\vartheta_0 < \frac{1}{2}\Bigg\{1-\frac{a_0^2}{\left(\frac{6 M a_0^2}{\left(\mu ^2+a_0^2\right)^{5/2}}-\frac{2}{a_0^2}\right)}\Bigg[& \frac{30 a^2_0 M}{\left(a^2_0+\mu ^2\right)^{5/2}}-\frac{4 M}{\left(a^2_0+\mu ^2\right)^{3/2}}-\frac{30 a^4_0 M}{\left(a^2_0+\mu^2\right)^{7/2}}-\frac{2 \Lambda }{3}\\
&-\frac{1}{2}\left(\frac{-\frac{4 M a_0}{\left(\mu ^2+a_0^2\right)^{3/2}}+\frac{6 M a_0^3}{\left(\mu ^2+a_0^2\right)^{5/2}}-\frac{2 \Lambda a_0}{3}}{1-\frac{2 M a_0^2}{\left(\mu ^2+a_0^2\right)^{3/2}}-\frac{\Lambda a_0^2}{3}}\right)^2\Bigg]\Bigg\}~.
\ee
Or
\be
\vartheta_0 < \frac{1}{2} \Bigg(1-\frac{-\frac{4 \Lambda a_0^2 \left(6 \mu ^2 M-3 M a_0^2+\Lambda  \left(\mu ^2+a_0^2\right)^{5/2}\right)^2}{3 \left(\mu ^2+a_0^2\right)^2 \left(6 M a_0^2+\left(\Lambda a_0^2-3\right) \left(\mu ^2+a_0^2\right)^{3/2}\right)^2}+\frac{30 M a_0^2}{\left(\mu ^2+a_0^2\right)^{5/2}}-\frac{4 M}{\left(\mu ^2+a_0^2\right)^{3/2}}-\frac{30 M a_0^4}{\left(\mu ^2+a_0^2\right)^{7/2}}}{\frac{6 M}{\left(\mu ^2+a_0^2\right)^{5/2}}-\frac{2}{a_0^4}}\Bigg)~.
\ee
\end{widetext}

%\begin{widetext}
\onecolumngrid
\vspace*{1cm}
%Graphs
\begin{figure}[!h]
%\centering
\begin{minipage}[t]{0.48\linewidth}
\captionsetup{justification=centering}
\includegraphics[width=\linewidth]{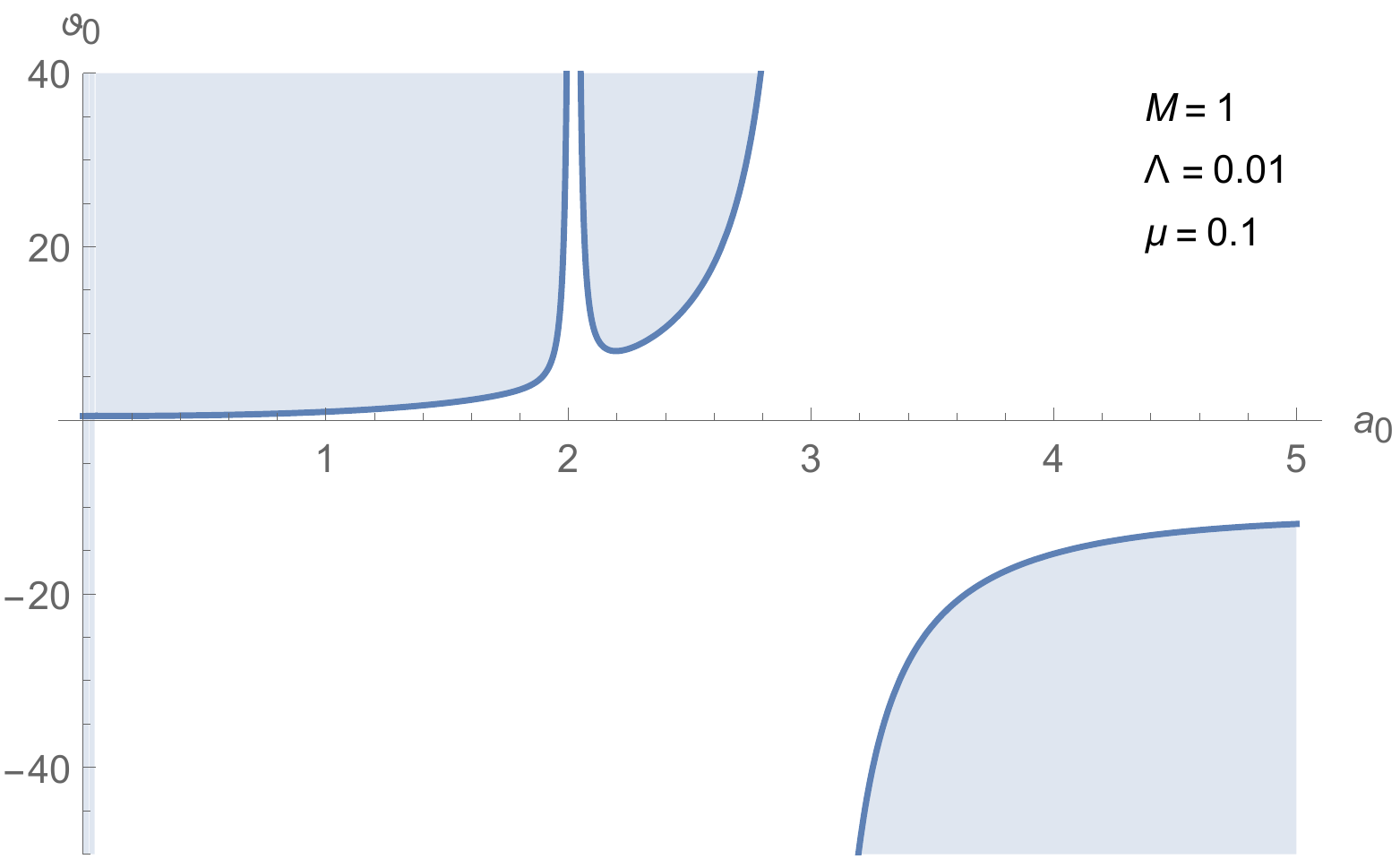}
\subcaption{ $\mu=0.1$.}
\end{minipage}\hfill
\begin{minipage}[t]{0.48\linewidth}
\captionsetup{justification=centering}
\includegraphics[width=\linewidth]{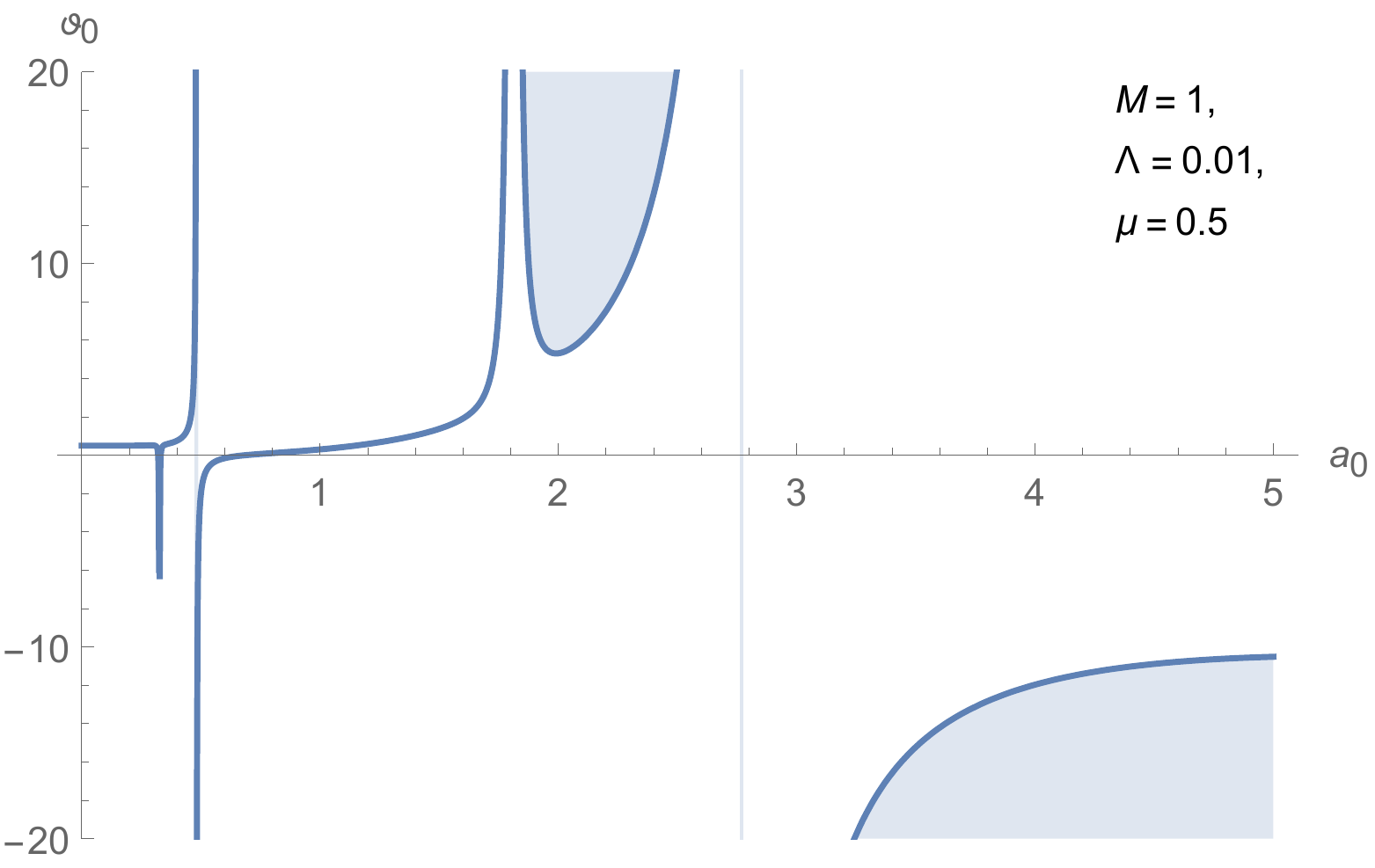}
\subcaption{ $\mu=0.5$.}
\end{minipage}

\begin{minipage}[t]{0.48\linewidth}
\captionsetup{justification=centering}
\includegraphics[width=\linewidth]{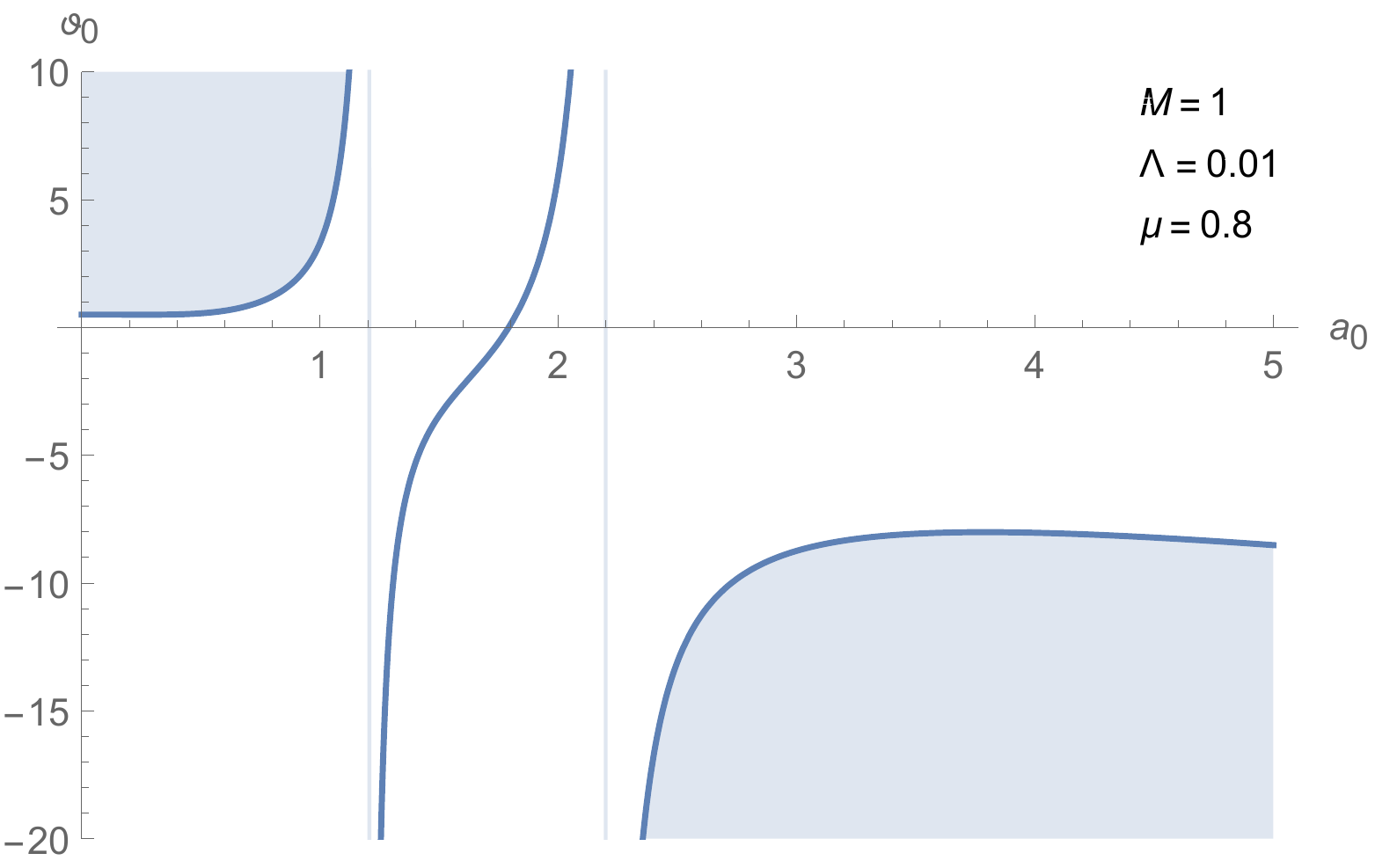}
\subcaption{ $\mu=0.8$.}
  \end{minipage}\hfill
\begin{minipage}[t]{0.48\linewidth}
\captionsetup{justification=centering}
\includegraphics[width=\linewidth]{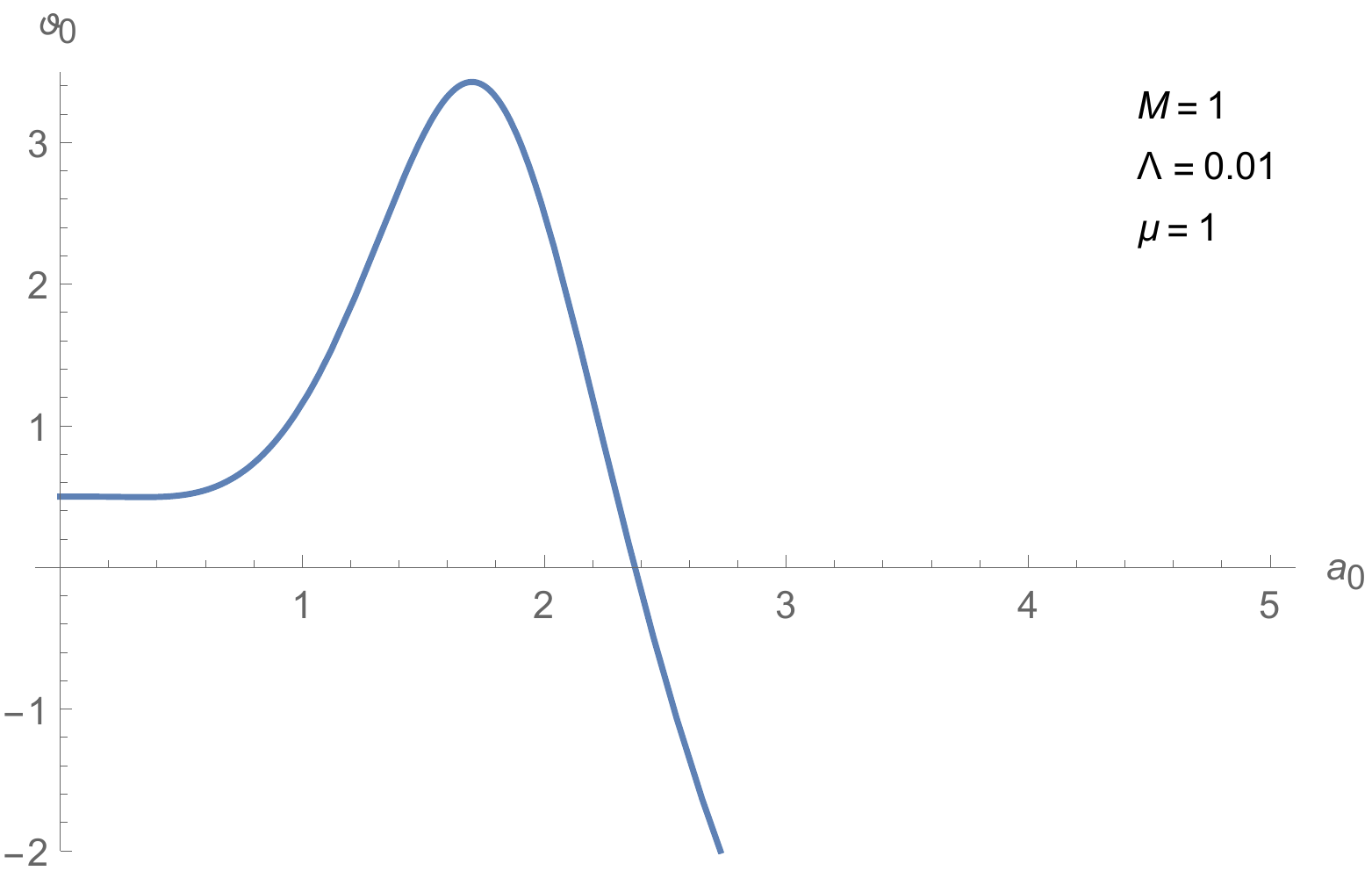}
\subcaption{ $\mu=1$.}
  \end{minipage}

\caption{Regions of stability of the thin-shell wormhole for the Bardeen de-Sitter solution for fixed values of $M=1$ and $\Lambda=0.01$, and different values of $\mu$. Stable regions are the blue shaded domains.}\label{fig.4}
\end{figure}

%\end{widetext}
\pagebreak
\twocolumngrid

%%%%%%%%%%%%%%
%
%
%%%%%%%%%%%%%%

\section{Discussion}
\par\noindent
In this letter we construct Bardeen de-Sitter thin-shell wormhole. We use Visser's technique of cut-and-paste with Darmois-Israel formalism to connect two BdS regions of spacetime through a thin shell. We compare the asymptotic behavior of the metric with that of SdS and RNdS as in fig.(\ref{fig.1}). We also study the components of the stress-energy-momentum surface tensor using the extrinsic curvature. We find that WEC is always violated. However, both NEC and SEC can be maintained upon imposing the inequalities that relate $f(r)$ to $f'(r)$. The energy conditions are shown in fig.(\ref{fig.2}). Then, we calculate the radial acceleration to express the attractive and repulsive nature of the wormhole throat. The results are plotted in fig.(\ref{fig.3}).\\
\par\noindent
Also we analyze the linear stability of BdS thin-shell wormhole by studying the concavity behavior on the ``speed of sound'' as a function in BdS parameters: the mass, the magnetic monopoles and the cosmological constant. And we see the change in stability regions upon varying the charge of magnetic monopoles while both mass and cosmological constant are fixed. The analysis is demonstrated in fig.(\ref{fig.4}). We conclude that for a diminutive value of cosmological constant and small value of magnetic charge, relative to the amount of mass, we find different regions of stability. Once the mass is equal to the magnetic charge, we no longer have stability regions. So to keep the Bardeen de-Sitter thin shell wormhole, and for a minute value of the cosmological constant, we suggest choosing the value of magnetic charge to be always less than the value of mass.
%%%%%%%%%%%%%%
\begin{center}
***
\end{center}
The author would like to thank the anonymous referee of the manuscript for the constructive suggestions to amend the presentation of the letter.
%%%%%%%%%%%%%%

%%%%%%%%%%%%%%
%
%

\end{arabicfootnotes}
\end{document}